\documentclass[aps,nobibnotes,prb,preprint,
nofootinbib,nobibnotes,lengthcheck]{revtex4-1}

\usepackage{hyperref}

\usepackage{amsmath,amssymb}
\usepackage{graphicx,xcolor}
\usepackage{bm,bbold}

\newcommand{\Gauss}{\;{\rm Gauss}\;}

\newcommand{\p}{\partial}
\newcommand{\id}[1]{\mathbb{1}_{#1\times #1}}

\newcommand{\Sgn}{\text{Sgn}\,}

\newcommand{\muR}{\mu_\textsc{r}}
\newcommand{\muL}{\mu_\textsc{l}}
\newcommand{\BAC}{B_\textsc{AC}}
\newcommand{\Prefactor}{\mathcal{C}_\textsc{cme}}
\newcommand{\epsilonF}{\varepsilon_\textsc{f}}
\newcommand{\epsilonNode}{\varepsilon_\textsc{w}}
\newcommand{\epsilonNodeR}{\varepsilon_\textsc{wr}}
\newcommand{\epsilonNodeL}{\varepsilon_\textsc{wl}}
\newcommand{\pF}{p_\textsc{f}}

\newcommand{\aNN}{a_\textsc{nn}}
\newcommand{\pzWR}{p^z_\textsc{wr}}
\newcommand{\pzWL}{p^z_\textsc{wl}}
\newcommand{\vid}{v_{\mathbb{1}}}
\newcommand{\epsilonDiag}{\varepsilon_\mathrm{diag}}

\newcommand{\lB}{l_\textsc{b}}

\newcommand{\muBohr}{\mu_\textsc{b}}

\newcommand{\Inversion}{\mathcal{P}}
\newcommand{\TimeReversal}{\mathcal{T}}

\begin{document}


\title{Non-universality of the adiabatic chiral magnetic effect\\ in a clean Weyl semimetal slab}



\author{Artem Ivashko}
\author{Vadim Cheianov}
\author{Jimmy A. Hutasoit}
\affiliation{Instituut-Lorentz for Theoretical Physics, Universiteit Leiden,
  P.O. Box 9506, 2300 RA Leiden, The Netherlands}
\affiliation{Delta Institute for Theoretical Physics, Science Park 904, 1090 GL Amsterdam, The Netherlands}


\date{\today}

\begin{abstract}
The adiabatic chiral magnetic effect (CME) is a phenomenon by which a slowly 
oscillating magnetic field applied to a conducting medium induces an electric current 
in the instantaneous direction of the field. Here we theoretically investigate 
the effect in a ballistic Weyl semimetal sample having the geometry of a slab. 
We discuss why in a general situation the  bulk and the boundary contributions towards the CME are comparable. We show, however, that
under certain conditions the adiabatic CME is dominated by 
the Fermi arc states at the boundary. We find that despite the topologically 
protected nature of the Fermi arcs, their contribution to the CME is neither 
related to any topological invariant nor can generally be calculated within the bulk low-energy effective 
theory framework. For certain types of boundary, however, the Fermi arcs contribution to the CME can be found from the effective low energy Weyl Hamiltonian and 
the scattering phase characterising the collision of a Weyl excitation with 
the boundary.
\end{abstract}


\maketitle

\section{Introduction}
Weyl semimetals (WSMs) are crystalline materials in which the low-energy electronic 
excitations are described by the Weyl Hamiltonian originating in the theory of 
massless relativistic fermions in four-dimensional space-time. 
Such materials were hypothesised more than three decades ago~\cite{nielsen1983adler}, 
then in the course of the last decade several chemical compounds were investigated
as candidates~\cite{murakami2007phase,wan2011topological,xu2011chern,burkov2011weyl,
weng2015weyl,huang2015weyl} culminating in 2015 in photoemission experiments 
showing quasi-Weyl dispersion of elementary excitations  
in a semi-metal~\cite{xu2015discovery,lv2015experimental,lv2015observation,yang2015weyl}
(see also Refs.~\onlinecite{hasan2017discovery} and~\onlinecite{armitage2017weyl} for recent reviews).
A typical WSM features an even number of singular points in its Brillouin zone 
in whose vicinity the effective single-particle Hamiltonian can be written as~\cite{hasan2017discovery,armitage2017weyl}
\begin{equation}
  \label{eq:linear-Hamiltonian-Weyl}
  \mathcal{H}_\text{eff} = \epsilonDiag(\bm p) \id{2} + \chi (\bm p - \bm p_0) \hat v \bm \sigma,
\end{equation}
where $ \bm p_0$ is the singular point called the ``Weyl node'', 
$\bm\sigma = (\sigma^x,\sigma^y,\sigma^z)$ is the ``pseudospin'', which does not necessarily coincide 
with the electron's spin, even though it behaves like a spin under the discrete spacetime symmetries and 
spatial rotations,  and $\epsilonDiag = \epsilonNode + \bm \vid (\bm p - \bm p_0)$ is a scalar
part of the energy. We assume that the tensor $\hat v$ is positive definite which enables us to introduce 
the chirality number $\chi=\pm 1$ characterising each Weyl node. 
We shall call the nodes having $\chi=1$ right-chiral and those having $\chi=-1$ left-chiral.

A characteristic macroscopic signature of the Weyl spectrum is the hypothetical 
Chiral Magnetic Effect (CME). The CME was originally predicted in 1980~\cite{vilenkin1980equilibrium} 
for ultra-relativistic plasmas, and later on it was discussed in the context 
of heavy-ion collisions~\cite{fukushima2008chiral,kharzeev2011testing,kharzeev2014chiral}, 
the early Universe~\cite{joyce1997primordial,boyarsky2012long,boyarsky2012self,
ivashkoCMEmass}, and relativistic magnetohydrodynamics in 
general~\cite{dvornikov2015magnetic,sigl2016chiral,
boyarsky2015magnetohydrodynamics}. In its simplest form, the CME 
is a phenomenon by which 
an electric current develops in the direction of a static magnetic 
field applied to a system in thermal equilibrium.
The CME requires that the system possesses 
an additional conserved parity-odd charge, which in the case of 
the Weyl Hamiltonian is the difference between the number of right-chiral 
and left-chiral particles. If the plasma is prepared in 
a thermal state such that the right-handed and the left-handed particles 
have different chemical potentials, $\muR$ and $\muL,$ then 
the application of a magnetic field $\mathbf B$ should result in the current 
density $\bm j = (\muR - \muL) \mathcal C \bm B $ where $\mathcal C = e^2/h^2c.$~\cite{kharzeev2014chiral}
(For simplicity, we restrict ourselves to a model with only two Weyl nodes throughout the paper.)

The newly discovered WSMs seem to be natural test beds for the observation of 
the CME. However such an experimental program is not without a problem. 
Indeed, in a realistic sample of a solid-state material the 
chirality quantum number is neither protected against impurity scattering
nor preserved in collisions with the sample boundary. Therefore continuous external 
driving is required in order to maintain the imbalance $\muR - \muL \neq 0$ 
~\cite{vazifeh2013electromagnetic,yamamoto2015generalized}.
One way to achieve this is to apply an electric field $\bm E$  parallel 
to the magnetic field, $\bm E \parallel \bm B$. In such a case, the mechanism 
responsible for the driving is the chiral 
anomaly~\cite{adler1969axial,bell1969pcac}, and it is believed to be the primary 
cause of the negative longitudinal magnetoresistance which is observed in 
transport experiments on 
WSMs~\cite{nielsen1983adler,son2013chiral,burkov2014chiral,li2016chiral,
zhang2016signatures}. It is worth noting, however, that the intrinsic effect of chiral 
anomaly can be masked by the other effects, e.g. related to the geometry of the 
measuring setup or the spatial variations of the sample conductivity, see Refs.~\onlinecite{dos2016search} and~\onlinecite{schumann2017negative}. Moreover, the negative longitudinal 
magnetoresistance was claimed to be observed in 3D materials \emph{without} any 
Weyl 
nodes~\cite{ganichev2001giant,wiedmann2016anisotropic,li2016negative,
li2016resistivity,liang2016pressure,luo2016anomalous,assaf2017negative}.

Another way to drive the system out of equilibrium is to make the magnetic field 
itself time-dependent, $\bm B(t) = \bm\BAC \cos \omega t$. Recent theoretical 
studies~\cite{chen2013axion,goswami2015optical,chang2015chiral,ma2015chiral,
zhong2016gyrotropic,alavirad2016role} converge in their conclusion that in 
a clean infinite sample such a perturbation will lead to the CME
of the form
\begin{equation}
  \label{eq:CME-AC}
  \bm j = \Prefactor \frac{e^2}{h^2c} b_0 ~ \bm\BAC \cos \omega t,
\end{equation}
where $b_0$ is the energy separation between 
the right-chiral and left-chiral Weyl nodes, 
$b_0 = \epsilonNodeR - \epsilonNodeL$. 
Note that the proportionality coefficient on the right-hand side 
of Eq.~\eqref{eq:CME-AC} is frequency-independent, therefore the formula 
predicts the effect in the adiabatic $\omega \to 0 $ limit. We shall call such 
a CME adiabatic. 

In a realistic sample the applicability of Eq.~\eqref{eq:CME-AC} 
is limited by a number of factors. Arguably, the most important one is 
the rate $\Gamma$ of chirality relaxation due to the impurity scattering.  
In the frequency range $\omega/\Gamma \lesssim 1$ chirality relaxation should 
dominate therefore the CME should be suppressed. Another, less obvious 
limiting factor is the geometry of the sample. Any physical sample has a finite 
cross-section and a boundary. Eq.~\eqref{eq:CME-AC} implies that the total 
CME current is proportional to the cross-sectional area $S_\perp$ of the sample and therefore, 
one may be tempted to think that the boundary effects would be irrelevant in samples with 
large cross-sectional areas. 
This turns out not to be the 
case~\cite{pesin2016nonlocal, baireuther2016scattering,ivashko2017adiabatic}.

In particular, the analysis of Ref.~\onlinecite{pesin2016nonlocal} exploiting general symmetry 
constraints on the structure of the gradient expansion of the polarisation tensor implies 
that the contribution of the boundary layer to the CME current 
is always one half of the bulk contribution no matter how big the sample. An alternative 
approach~\cite{baireuther2016scattering} based on microscopic analysis for a particular model of 
WSM arrives at a similar conclusion: the boundary contribution to the CME current is on the same order as 
the bulk contribution albeit the numerical coefficient is two rather than one half. 
These two results are quite remarkable in both their agreement as to the scale 
of the boundary effect, and their disagreement in regards to the numerical factor
defining the actual value of the boundary current relative to the bulk.
What is the reason for the discrepancy?
The gradient expansion of kinetic coefficients used in Ref.~\onlinecite{pesin2016nonlocal}
implicitly assumes that these coefficients are (quasi)local.
For ballistic systems, however, the low-frequency response is known to be
highly non-local, which can be seen already from the fact that the limits
$\omega \to 0$ and $k \to 0$ do not commute, for an unbounded 
sample~\cite{ma2015chiral,zhong2016gyrotropic,pesin2016nonlocal}.
($k$ here is the wavevector of the magnetic field, for more details
about the non-locality, see Ref.~\onlinecite{ivashko2017adiabatic}.) For the gradient 
expansion to work in a finite-size sample the frequency of the magnetic 
field has to be much greater than $v/L$ where $v$ is the typical 
speed of an elementary excitation and $L$ is the typical size of the sample's 
cross-section. In contrast, the approach of Ref.~\onlinecite{baireuther2016scattering} 
is valid in the opposite low-frequency (adiabatic) limit~\cite{ivashko2017adiabatic} outside 
the applicability range of the gradient expansion theory. In the present paper, we further
investigate the boundary contribution to the CME current in the adiabatic limit 
in order to address the following questions a) Is the coefficient ${\mathcal C}_B=1$ in the 
boundary current $I_B ={\mathcal  C}_B (b_0 e^2/h^2c) S_\perp B$ found in Ref.~\onlinecite{baireuther2016scattering} universal
(possibly topologically protected)? 
b) If it is not, can it be nevertheless expressed in terms of the parameters of the effective 
low-energy theory including the Weyl Hamiltonian of the elementary excitations and 
the scattering matrix at the boundary?
Our main finding is that the answer to both questions is generally ``no'' although 
under certain conditions the answer to question b) can be positive. 

The paper is organized as follows. In Sec.~\ref{sec:background} we discuss the 
methods that we use for the analysis of the adiabatic CME, and the particular 
set-up. In Secs.~\ref{sec:bulk-effective-theory} and~\ref{sec:surface-effective-theory} we discuss the contributions 
of the bulk and the boundary to the adiabatic CME in the framework of effective 
low-energy theory. In Sec.~\ref{sec:surface-microscopic-theory}, we take into 
account the contribution of boundary that is not captured by the effective 
theory, by using the same microscopic model as in~\cite{baireuther2016scattering}. In 
Sec.~\ref{sec:discussion}, we discuss our findings.

\section{Methods and setup}
\label{sec:background}
For definiteness, we consider a sample having the geometry of a slab which is 
infinite in the $y-z$ plane and has   
thickness $L_\perp$ in the $x$ direction. 
We assume that the sample is in the state of thermal 
equilibrium at temperature $T=0,$ and we denote the Fermi energy $\epsilonF.$ The 
oscillating magnetic field is applied along the $z$-axis.

It is worth noting that we consider a sample geometry which 
is slightly different from the geometry of an infinite cylinder with 
a compact base investigated in Ref.~\onlinecite{baireuther2016scattering}. The 
original choice of Ref.~\onlinecite{baireuther2016scattering} was motivated 
by the considerations of numerical convenience in application of the following 
heuristic formula for the total electric current $I$ along the cylinder's axis
\begin{equation}
  \label{eq:adiabatic-Kubo-current}
 I = 
 \frac{e \BAC}{h} \sum\limits_\nu \int\limits_\text{BZ} dp ~ \theta(\epsilonF-\varepsilon_\nu(p)) \frac{\p^2 \varepsilon_\nu(p)}{\p B \p p}.
\end{equation}
Here $p$ is the quasimomentum along the magnetic field, which runs over
the one-dimensional Brillouin zone (BZ) of the cylinder, $\nu$ is an
additional index that characterizes the energy levels. In the
recent paper~\onlinecite{ivashko2017adiabatic},
Eq.~\eqref{eq:adiabatic-Kubo-current} was derived from the first-principle 
quantum-mechanical linear-response theory, where it was shown that the formula is
applicable only for the \emph{adiabatic} driving, meaning that the driving
frequency $\omega$ is much less than the spacing between any pair of energy
levels associated with a non-vanishing matrix element of the velocity or the magnetic moment operators.

For the slab geometry considered here, the index $\nu$ comprises the quasimomentum 
$p^y$ along the $y$-axis and some additional discrete index $n.$ In this case 
the relevant matrix elements between the states having either different $p^y$ or different $p$ 
vanish due to the translational invariance in $y$ and $z$-directions. As a 
result, adiabaticity can be broken only in transitions between different $n$. 
Note that in the limit of large thickness, $L_\perp \to \infty$ the level spacing between 
the states of different $n$ collapses, which leads to the breakdown of adiabaticity. One of the 
ways to restore adiabaticity in such a limit is to apply a large  
\emph{static} background magnetic field $B_0$, which we choose to be 
directed along the $z$-axis, such that the total field is $B = B_0 + \BAC \cos \omega 
t$. While the bulk Landau levels are separated by finite 
energy gaps on the order $v \sqrt{eB_0 \hbar/c}$ (see 
Sec.~\ref{sec:surface-effective-theory} for more details), it is in 
principle possible that for some surface states there is one or more pair of levels 
with a significantly smaller energy spacing. However, we expect 
this to occur very rarely as we change $p^y$ for a 
fixed $p$, since at the same time, these pairs of states must be close to the 
Fermi energy, in order to contribute to the 
current~\eqref{eq:adiabatic-Kubo-current}. (This expectation of the rare 
crossings is confirmed by the numerical  
calculations for a particular model used below.) 

The adiabatic regime has an obvious advantage from both analytical and numerical
points of view. Namely, in order to find the current $I$ it is enough to
know the single-particle energy spectrum $\varepsilon_\nu(p)$, while in the
non-adiabatic regime we need to calculate additionally the off-diagonal
matrix elements of the velocity and the magnetic moment
operators~\cite{ma2015chiral,ivashko2017adiabatic}.

In order to separate the bulk and the surface components of the current from 
Eq.~\eqref{eq:adiabatic-Kubo-current}, we use the result 
of Ref.~\onlinecite{ivashko2017adiabatic}~~\footnote{We note that 
in Ref.~\onlinecite{ivashko2017adiabatic}, the derivation was done for the geometry of a 
cylinder with a \emph{circular} base. However, the generalization to the 
geometry of a slab is straightforward: one only needs to replace the momentum 
$p_\parallel$ corresponding to the motion along the perimeter of the circle with 
the momentum $p^y$, and to take into account that cylinder has only one 
boundary, while the slab has two. Additional factor $1/2$ in 
Eq.~\eqref{eq:surface-current-slab} is due to the fact that the inflow of the 
charge from the bulk to the boundary is splitted between the two boundaries. 
(For more details about the inflow mechanism, see Ref.~\onlinecite{ivashko2017adiabatic}.)}
The surface current is found from the following formula
\begin{equation}
  \label{eq:surface-current-slab}
  I_\text{surf} = \BAC \frac{e^2}{2h^2 c} S_\perp \int\limits_\text{BZ} dp 
  \sum\limits_{n,\pm} \left( v_n^z ~ \Sgn \frac{\p \varepsilon_n}{\p p^y} \right) \Big|_{\varepsilon_n = \epsilonF} \rho_n(p),
\end{equation}
where $S_\perp = L_\perp L_y$ is the area of the sample cross-section, $v_n^z = 
\p \varepsilon_n/ \p p$ is the group velocity along the magnetic field, the sum 
goes over the states localized at the right ($+$) and the left ($-$) boundaries, 
$\rho_n(p) = 1$ if there exists a solution of $\varepsilon_n(p,p^y) = \epsilonF$ 
for given $n$ and $p$, and $\rho_n(p) = 0$ otherwise. (In order to deal with 
finite $S_\perp$, we have introduced a finite width $L_y$ in the $y$-direction, 
but we assume that this width is much larger that any other length scales 
in our problem.) The bulk current is given by the expression
\begin{equation}
  \label{eq:bulk-current}
  I_\text{bulk} = \frac{e^2 \BAC}{h^2 c} S_\perp \sum\limits_n \sum\limits_{\pF} \frac{\p \varepsilon_n(\pF)}{\p B} \Sgn v^z_n(\pF),
\end{equation}
where $\varepsilon_n(p)$ is the energy of the bulk levels, and we drop $p^y$ here 
owing to the fact that the bulk energy levels (Landau levels) are degenerate 
with respect to this quasimomentum. The sum goes over all solutions of the equation 
$\varepsilon_n(\pF) = \epsilonF$. Note that both $I_\text{bulk}$ and 
$I_\text{surf}$ scale linearly with the area $S_\perp$.

Note that the surface CME contribution is different from the well-studied dia- 
or para-magnetic surface currents. First, the latter appear even in thermal 
equilibrium, in the absence of a time-dependent component $\delta \bm B$. Second, 
the total current through the cross-section calculated from dia-/para-magnetic 
current density $\bm j_\text{eq} = c ~ \bm\nabla \times \bm (\hat 
\chi_\text{magn} \bm B)$ is zero. (Here $\hat \chi_\text{magn}$ is the magnetic 
susceptibility tensor.)

For our numerical analysis in Sec.~\eqref{sec:surface-microscopic-theory}, we employ the 
same microscopic model that was used in Ref.~\onlinecite{baireuther2016scattering}, which 
is a four-band tight-binding model with the single-particle Hamiltonian
\begin{equation}
  \label{eq:Vazifeh-Franz-Hamiltonian}
  \mathcal{H}_\text{lattice} = \begin{pmatrix} \mathcal{H}_{11} & \mathcal{H}_{12} \\ \mathcal{H}_{12}^\dagger & \mathcal{H}_{22} \end{pmatrix},
\end{equation}
where
\begin{eqnarray}
  \mathcal{H}_{11} &=& 2t(\sigma^x \sin p^x + \sigma^y \sin p^y) + \frac{\beta^z}{2} \sigma^z, \\
  \mathcal{H}_{22} &=& - 2t(\sigma^x \sin p^x + \sigma^y \sin p^y) + \frac{\beta^z}{2} \sigma^z, \\
  \mathcal{H}_{12} &=& - i t \sin p^z + M(\bm p) - i \frac{\beta_0}{2} \sigma^z, \\
  M(\bm p) &=& M_0 + t (3 - \cos p^x - \cos p^y - \cos p^z).
\end{eqnarray}
Here, $t$ describes the nearest-neighbour hopping, $\beta_0$ and $\beta_z$
are parameters that violate the inversion $\Inversion$ and time-reversal $\TimeReversal$ symmetries,
respectively, $\sigma^x, \sigma^y,$ and $\sigma^z$ are the pseudospin
operators, $\bm p$ is the quasimomentum. Breaking $\Inversion$ is required in order to have non-vanishing difference of energies $b_0 = \epsilonNodeR - \epsilonNodeL$, and we are forced to break the time-reversal symmetry in order to deal with only two Weyl nodes. (The minimal number of nodes in presence of $\TimeReversal$ is four~\cite{armitage2017weyl}.) The lattice has cubic unit cell,
and for simplicity, we take the lattice spacing equal to 1,
so that $\bm p$ is measured in units of $\hbar$.

\section{Adiabatic bulk CME in the effective theory}
\label{sec:bulk-effective-theory}
In this Section, we study the adiabatic CME current in the framework of effective theory. 
First, we recall that in an idealised model of a Weyl semimetal neglecting both 
the momentum dependence of the scalar part $\epsilonDiag(\bm p)$
of the effective Hamiltonian~\eqref{eq:linear-Hamiltonian-Weyl} and the gradient 
corrections to the linear spectrum, the bulk contribution to the CME is suppressed 
in the presence of a background magnetic field $B_0$~\cite{ivashko2017adiabatic}.
This can be seen from inspecting the dispersion relations of Landau 
levels~\cite{johnson1949motion,berestetskii1982quantum} that 
enter Eq.~\eqref{eq:bulk-current},
\begin{equation}
  \label{eq:bulk-Landau-level-dispersion}
  \varepsilon_n - \epsilonNode = \begin{cases} -\chi v^z \delta p, & (n = 0) \\ \Sgn n \cdot \sqrt{v_z^2 \delta p^2 + 2 |n v^x v^y|\frac{\hbar^2}{\lB^2}}. & (n \neq 0) \end{cases}
\end{equation}
The index $n$ here is the number of the Landau level, which is
  the effective low-energy counterpart of the index $n$ introduced earlier in this paper,
$\delta p = p - p^z_0,$  $\lB = \sqrt{eB_0/\hbar c}$ is the magnetic length. 
Since the energy of the $n=0$ level does not depend on the magnetic field,
this level does not contribute to the current $I_\text{bulk}$,
according to Eq.~\eqref{eq:bulk-current}. Although 
the spectrum of the $n \neq 0$ levels involves the magnetic field,
their energies are \emph{even} with respect to the difference $p - p^z_0$, 
which means that they do not contribute to the bulk current either. 

The vanishing of the bulk CME in a simplified model is accidental and it 
is not protected against various deformations of the Hamiltonian.
We identify the following main factors that might lead to a non-vanishing bulk 
current in a more realistic model.
Firstly, Eq.~\eqref{eq:bulk-Landau-level-dispersion} 
is only valid if the Landau quantization of energy levels is stronger
than the finite-size quantization. This implies that the magnetic field $B_0$ 
needs to be strong enough to ensure the condition $\lB \ll L_\perp$.
For weak background magnetic field violating this bound, 
the structure of energy levels becomes different, and an appreciable bulk current 
may develop, in agreement with Ref.~\onlinecite{baireuther2016scattering}. Secondly,
the minimal effective Hamiltonian~\eqref{eq:linear-Hamiltonian-Weyl} 
is applicable only in the long wavelength limit. 
Gradient corrections to this Hamiltonian will generally modify the dispersion 
relations in a way which will lead to a finite bulk CME current. 
We discuss such corrections in App.~\ref{app:higher-order-corrections}, and present arguments as 
to why they are negligible under realistic assumptions. Finally, a violation of 
the assumption $\bm\vid = \bm 0$ may also lead to a bulk current within the 
chosen model. This can be easily seen, for instance, in the situation of 
$\bm \vid \parallel \bm B.$ It can, however, be shown that in this situation 
non-vanishing contributions from different Weyl points cancel if the 
system as a whole possesses time-reversal symmetry. Our reason for working with 
a model breaking time-reversal symmetry is that we want to compare our 
results with Ref.~\onlinecite{baireuther2016scattering}.
It is purely accidental that in the parametric range used 
in Ref.~\onlinecite{baireuther2016scattering} the effective parameter $\bm \vid$ 
turns out to be negligible thus emulating the effect of the time-reversal symmetry 
protected cancellation.

To conclude, the bulk adiabatic CME current is not related to the 
oscillating magnetic field in a universal way. Its material part, however, is 
small and its geometric part is controllable and can in principle be tuned to vanish.

\section{Adiabatic boundary CME in the effective theory}
\label{sec:surface-effective-theory}
Next, we turn to the analysis of the surface contribution~\eqref{eq:surface-current-slab} 
to the adiabatic CME trying to approach the problem from the bulk low-energy effective theory perspective. 
In order to describe a bounded system, one has to supplement the effective
Hamiltonian in Eq.~\eqref{eq:linear-Hamiltonian-Weyl} with the boundary condition
on the single-particle wavefunction $\psi$. The generic condition
for a boundary located at $x = x_\textsc{b}$ has the form~\cite{witten2015three}
\begin{equation}
  \label{eq:boundary-condition-Weyl}
  (\sigma^y \sin\Delta\phi + \sigma^z \cos\Delta\phi + 1) \psi\Big|_{x=x_\textsc{b}} = 0,
\end{equation}
where $\Delta\phi$ has the meaning of the scattering 
phase shift at the surface.
The effective low-energy theory offers no constraints on the parameter $\Delta \phi.$
The actual value of the phase shift depends on the microscopic detail of the boundary.
In a sample having two boundaries, $x = \pm L_\perp/2$, each boundary is 
characterised by the condition~\eqref{eq:boundary-condition-Weyl} with its own 
value of $\Delta\phi$. 
Moreover, each Weyl node has its own scattering phase shift, which means that in 
our particular setup we have four independent phase shifts in total. 
In the rest of the text, we denote them as $\Delta \phi^\pm_\textsc{l,r}$, 
where the upper index corresponds to the $x = \pm L_\perp/2$ boundaries, while L (R) denotes the left- 
(right-)chiral node.

In the presence of a constant background field $B_0$ the eigenvalues of the system's Hamiltonian 
can be classified by the $z$-projection of quasi-momentum $p$ and the eigenvalue $p^y$ of the 
operator of magnetic translations in the $y$-direction. As is usual in the theory of Landau quantization~\cite{brown1964bloch}, 
an orbital characterised by a given $p^y$ is localised within a distance $\lB$ from the plane 
$x=p^y \lB^2/\hbar.$
The equation which defines the dispersion relation 
$\varepsilon = \varepsilon_n(p,p^y)$ for the effective 
Hamiltonian~\eqref{eq:linear-Hamiltonian-Weyl} in the presence of a magnetic field and 
the boundary conditions~\eqref{eq:boundary-condition-Weyl} is quite cumbersome. 
However, in the limiting case we are interested in, $\lB \ll L_\perp$, 
it can be simplified to the following form
\begin{multline}
  \label{eq:boundary-state-spectrum-magnetic-field}
  (\varepsilon - \epsilonNode + \chi v^z \delta p) D_{\lambda/2 -1}(\tilde p^y) = \\ = \pm 
  \sqrt{2} \chi \frac{v_\perp}{\lB} \tan\left( \frac{\Delta\phi^\pm}{2}\right) D_{\lambda/2}(\tilde p^y),
\end{multline}
where the $\pm$ index is chosen depending on whether the orbital is localised 
near the $x=L_\perp/2$ or $x=-L_\perp/2$ boundary. Here, we have used the following notation:
$\lambda \equiv \left((\varepsilon - \epsilonNode)^2 - 
  v_z^2 \delta p^2 \right) \lB^2 / v_\perp^2$,
$\tilde p^y \equiv \sqrt{2} p^y \lB / \hbar \pm L_\perp /\sqrt{2} \lB$, and
$D_\nu(x)$ is the parabolic cylinder 
function~\cite{NIST:DLMF}. We have also chosen $v^x = v^y = v_\perp$.

\begin{figure}
  \centering
  \includegraphics[width=0.45\textwidth]{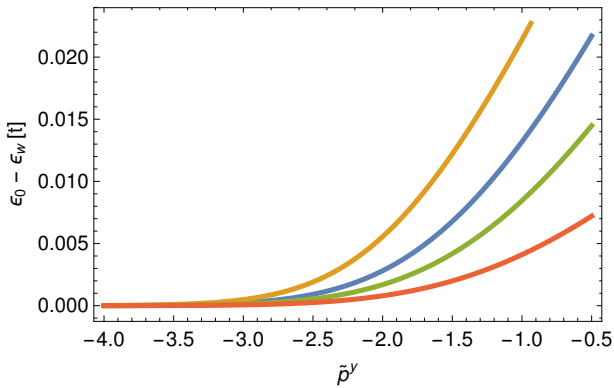}
  \caption{Dependence of the energy $\varepsilon_{n=0}$ on $\tilde p^y$ for a right-chiral electron in strong static magnetic field $B_0$,
    computed from Eq.~\eqref{eq:boundary-state-spectrum-magnetic-field}.
    From top to bottom: $\Delta\phi_\textsc{r}^- = -2.3, -1.49, -0.7$, and $-0.2$. $p = p^z_0$, while the other
    parameters of the effective theory are the same as in Eq.~\eqref{eq:effective-parameters}.
    The states with $\tilde p^y \lesssim -3.5$ are completely localized within the bulk
    (Landau levels), the other states are localized at the boundary $x = - L_\perp/2$.}
  \label{fig:boundary-states-dispersion}
\end{figure}

Note that
Eq.~\eqref{eq:boundary-state-spectrum-magnetic-field} encodes the dispersion 
relation of \emph{both} bulk and surface modes. For the states that
are localized at the surface (within the length $\delta x \ll \lB$), in the absence
of the magnetic field, turning on magnetic field does not affect the dispersion relation much.
Generally, in a Weyl semimetal,
there is at least one family of such surface states at the Fermi energy
called the Fermi arc~\cite{hasan2017discovery}. By tuning $p^y$ at fixed $p$, such a surface branch continuously 
transforms into one of the bulk Landau levels showing no energy dependence on $p^y.$ This behaviour is
illustrated in Fig.~\ref{fig:boundary-states-dispersion}, where
Eq.~\eqref{eq:boundary-state-spectrum-magnetic-field} is solved numerically
for the $n=0$ state. 

It was discussed in Ref.~\onlinecite{ivashko2017adiabatic} that the surface
contribution~\eqref{eq:surface-current-slab} to the adiabatic CME
originates in the inflow of electric charge from the bulk to the
boundary, which for every given $p$ is similar to the Hall effect arising in two-dimensional
systems~\cite{girvin9907002quantum}. In this picture, the chirality 
of the edge mode, which is defined as the sign of $\p \varepsilon_n / \p p^y$
at Fermi level, is linked to a topologically
protected characteristic (Chern number) of the Weyl semimetal~\cite{armitage2017weyl}
in that both describe the direction of
the inflow of charge (towards the boundary or away from it) at given value of $p.$ 
It is for this reason
that the chirality of each edge mode enters as a multiplier in Eq.~\eqref{eq:surface-current-slab}. 

It follows from the above topological considerations that the support function $\rho_n(p)$ in 
Eq.~\eqref{eq:surface-current-slab} is non-vanishing as long as the momentum $p$ is
inside the region of the one-dimensional projection of the Brillouin 
zone where the Chern number is finite (this region is approximately bounded by the 
positions of the two Weyl points). A significant part of this region lies outside 
the applicability range of the Weyl Hamiltonian, therefore there is no general 
reason to believe that the integral on the right-hand side of Eq.~\eqref{eq:surface-current-slab} 
can be calculated from the parameters of the effective low-energy theory. 
There still exists one noteworthy exception which is when the dispersion
relation of Fermi arcs is separable, meaning that in the absence of magnetic field
\begin{equation}
  \label{eq:FA-separability}
  \varepsilon_0(p,p^y) = \mathcal{G}_1(p) + \mathcal{G}_2(p^y),
\end{equation}
where  $\mathcal{G}_{1,2}$ are some arbitrary functions.
Here and below we choose the index $n=0$ for the branch of topologically non-trivial surface states.
Indeed for $B_0 = 0$, the contribution of the surface states is insensitive to 
what happens outside the vicinity of the Weyl nodes. This is because
according to Eq.~\eqref{eq:surface-current-slab}, we get the 
integral of total derivative, which reduces to the difference of the boundary 
values of the function $\mathcal{G}_1$ at the points $p = \pzWR$ and $p = 
\pzWL$. These values themselves are fixed by the positions of 
the energies of the Weyl nodes $\epsilonNodeR$ and 
$\epsilonNodeL$, respectively, so that the partial contribution of the Fermi arc is
\begin{equation}
  \label{eq:Fermi-arc-current}
  I^\textsc{fa}_\text{surf}(B_0=0) = \frac{e^2}{h^2 c} \BAC S_\perp (\epsilonNodeR - \epsilonNodeL).
\end{equation}
This is in agreement with Ref.~\onlinecite{baireuther2016scattering}, where the contribution of these surface states to the total coefficient
  $\Prefactor$ is argued to be equal to 1.

In case of non-vanishing $B_0$, the 
contribution to Eq.~\eqref{eq:surface-current-slab} beyond the effective theory 
remains unchanged, therefore the integral again reduces to the contribution 
from the states at the vicinity of the Weyl nodes. The latter is modified by 
the magnetic field albeit in a way that is completely defined by the low-energy effective 
theory. In order to specify Eq.~\eqref{eq:surface-current-slab} for this case,
we note that within the effective theory, in the absence of the magnetic field, the energy of the Fermi arc is linear with respect
  to both $p$ and $p^y$~\cite{witten2015three}. It
  means that even in the case of $B_0 \neq 0$, the separability holds both
  in the effective theory and in the ultraviolet complete theory, provided that
  the deviation from the Weyl node is large enough, $\delta p \gtrsim \Lambda_0$.
  Therefore, one can artificially split the integration region in Eq.~\eqref{eq:surface-current-slab} into two parts:
  the vicinities of the Weyl nodes, $\delta p \leq \Lambda$ ($\Lambda \gtrsim \Lambda_0$),
  and the rest. For the vicinities of the two Weyl nodes, we can still use the
  effective theory and we denote the result as $I_\text{eff}(\Lambda)$. Due to
  the linear dispersion, this contribution grows linearly with the cutoff $\Lambda$.
  In the remaining region, the separable relation~\eqref{eq:FA-separability} holds.
  The corresponding contribution
  is a linear function of $\Lambda$, which is equal to~\eqref{eq:Fermi-arc-current} at $\Lambda = 0$,
  and its slope is opposite to that of $I_\text{eff}$.
  Since the total surface CME current is not sensitive to the choice of the cutoff $\Lambda$, we can formally set it to infinity:
\begin{equation}
  \label{eq:surface-current-separable}
  I_\text{surf} = \lim\limits_{\Lambda \to \infty} \left( I_\text{eff}(\Lambda) - \Lambda \frac{d I_\text{eff}(\Lambda)}{d\Lambda} \right) + I^\textsc{fa}_\text{surf}(B_0=0).
\end{equation}

For the actual calculation using Eq.~\eqref{eq:surface-current-separable} we have used the effective parameters of the bulk Hamiltonian
\begin{eqnarray}
  \label{eq:effective-parameters}
  \epsilonNodeR = -\epsilonNodeL = 4.9 \times 
  10^{-2} t, \quad \epsilonF = 0, \\ v^z = 0.69 t/\hbar, \quad v_\perp = 2.0 t/\hbar, \quad \lB = 50,
\end{eqnarray}
and the two Weyl nodes share the same values of $v^z$ and $v_\perp$. 
Here $t$ is an arbitrary parameter that has dimension of energy. The 
specific choice of all these input parameters was done in order to compare
them 
(see below) with the results of a particular microscopic calculation and with the 
results of Ref.~\onlinecite{baireuther2016scattering}. We have found that even in the case
of separable Fermi arcs, the resulting prefactor in~\eqref{eq:CME-AC} depends
on the choice of $\Delta \phi$: for $\Delta \phi^\pm_\textsc{r} = -\Delta \phi^\pm_\textsc{l} = \pm 1.49$,
$\Prefactor = 0.53$, while for $\Delta \phi^\pm_\textsc{r} = -\Delta \phi^\pm_\textsc{l} = \pm 2.3$, the result changes to $\Prefactor = 0.89$.
(As before, the $+$ index corresponds to the boundary $x = L_\perp/2$, while the $-$ index corresponds to
the other boundary.) 

To conclude, the $\Prefactor$ is not universal in the sense that it depends on both the bulk and the \emph{boundary} parameters of the effective theory, \emph{even} in the case of separable energy of Fermi arcs. As we have 
noted above, if the separability does not hold, the resulting current $I_\text{surf}$ 
involves microscopic details of the material beyond the information encoded in 
the parameters of the effective theory. Therefore, in order to understand 
whether the separability is a generic property of a WSM, in the following Section we study a 
microscopic model of such a material.

\section{Surface contribution in a microscopic theory}
\label{sec:surface-microscopic-theory}
As we have established in the previous sections, the boundary CME is a significant 
effect, which under certain conditions dominates the longitudinal response of a WSM 
to the applied magnetic field. Although the Fermi arc states responsible for the effect
are topologically protected, the magnitude of the boundary CME cannot be linked to 
any topological invariant. Moreover, the magnitude of the 
boundary CME is generally not fully determined by the parameters of the low-energy effective theory in 
the bulk material unless a special condition on the dispersion relation of the Fermi 
arc states is met. There is no obvious reason why this condition should hold for an arbitrary 
material interface therefore it is natural to assume that it is likely to be violated in a given 
experimental sample. Nevertheless, it is instructive to see how the condition 
breaks down in a particular microscopic model and what consequences this may have 
for the boundary CME. 

To this end, we turn to the microscopic lattice Hamiltonian~\eqref{eq:Vazifeh-Franz-Hamiltonian}. 
The Bloch spectrum of the model contains one right-handed Weyl node located at
$\bm p_0(\text{right}) = (0,0,\pzWR)$, and one left-handed Weyl node located at
$\bm p_0(\text{left}) = - \bm p_0(\text{right}).$
The Bloch momentum $\pzWR$ can be expressed in terms of the parameters of the Hamiltonian as described in
Appendix~\ref{app:appendix}. By performing a unitary
transformation from the original basis to the eigenbasis of 
the Hamiltonian at $\bm p = \bm p_0(\text{right})$, together with the linearisation with respect to $\bm p - \bm p_0(\text{right})$, 
the resulting effective Hamiltonian is
\begin{equation}
  \label{eq:right-chiral-effective-Hamiltonian-VF}
  \mathcal{H}_\text{R} = \epsilonDiag (p^z) \id{2} + v_\perp(p^x \sigma^x + p^y \sigma^y) + v^z (p^z - \pzWR) \sigma^z.
\end{equation}
Here $\epsilonDiag(p^z) = \epsilonNodeR + \vid (p^z - \pzWR)$ and the relationships 
between the effective parameters and the parameters of the microscopic theory~\eqref{eq:Vazifeh-Franz-Hamiltonian}
are listed in Appendix~\ref{app:appendix}. The 
Hamiltonian~\eqref{eq:right-chiral-effective-Hamiltonian-VF} acts on the 
two-dimensional space which corresponds to the two gapless branches of the 
Hamiltonian's Bloch spectrum near the Weyl point. The two remaining 
gapped branches have been dropped from 
the effective theory. Note that 
Eq.~\eqref{eq:right-chiral-effective-Hamiltonian-VF} has the same form as the 
effective Hamiltonian \eqref{eq:linear-Hamiltonian-Weyl}.

For the \emph{left}-chiral node, a similar procedure leads to the same 
effective 
Hamiltonian~\eqref{eq:right-chiral-effective-Hamiltonian-VF}, but with
the replacement $(\pzWR,v^z,\epsilonNodeR) \to (-\pzWR,-v^z,-\epsilonNodeR)$.
Note that 
the energies of the two Weyl nodes are equal in magnitude and have 
opposite signs, therefore the  ``symmetric'' choice of the Fermi energy $\epsilonF = 0$ 
results in equal density of oppositely charged carriers in the two Weyl pockets. 
For simplicity and in conformity with the reference study~\onlinecite{baireuther2016scattering},
we limit our considerations to this symmetric situation.

For the actual numerical computations we have used one of the parameter sets from Ref.~\onlinecite{baireuther2016scattering}, namely $\beta_0 = 
0.1t$, $\beta^z = 1.2t$, $M_0 = -0.3t$, and $\lB = 50$. By using the dictionary in 
Appendix~\ref{app:appendix}, one can see that this choice leads to the
effective paramters listed in Eq.~\eqref{eq:effective-parameters}. The
smallness of $\epsilonNodeR - \epsilonF$ ensures that the Fermi surfaces
of the bulk states are close to the Weyl nodes, 
so that the effective theory describes the dynamics adequately. 
In addition, the parameter $\vid^z$ turns out to be very small
($\vid^z = -1.1 \times 10^{-2} t / \hbar$), which results in 
accidental suppression of $I_\text{bulk}$ in a strong background field $B_0$ 
(see Sec.~\ref{sec:bulk-effective-theory}).

We compute the energy spectrum of the bounded system,
using the \texttt{KWANT}
package~\cite{groth2014kwant}. This then serves as the input for the
equation~\eqref{eq:surface-current-slab}.
Since we deal with a sample that is 
infinite in both $y$- and $z$-directions, we perform the dimensional reduction from 
three to one dimensions by replacing the operators $p$ and $p^y$ with 
corresponding good quantum numbers.
We include the magnetic field by employing the standard Peierls substitution $\bm p 
\to \bm 
p - e \bm A / c$, and we choose the Landau gauge $\bm A = (0,Bx,0)$, which does 
not break the one-dimensional character of the problem. We have used 800 
lattice sites in the $x$-direction, and checked that the doubling of this 
number does not change significantly any of the results presented below.

We have checked that the energy dispersion~\eqref{eq:bulk-Landau-level-dispersion} 
for the Landau levels holds well and the deviations (beyond having non-zero $\vid^z$) are in 
agreement with the dimensional analysis done in Appendix~\ref{app:higher-order-corrections}.

We confirm that the boundary condition for the low-energy excitations that follows from the microscopic 
Hamiltonian~\eqref{eq:Vazifeh-Franz-Hamiltonian}
with the hard-wall boundaries indeed has
the form~\eqref{eq:boundary-condition-Weyl}.
The extracted scattering phase-shifts turn out to be
$\Delta\phi^\pm_\textsc{r} = -\Delta\phi^\pm_\textsc{l} = \pm 1.49$, which
coincide with one of the two combinations that we used in Sec.~\ref{sec:surface-effective-theory}.
Futhermore, the numerical evaluation of the surface
current~\eqref{eq:surface-current-slab} using \texttt{KWANT} gives the
\emph{same} value $\Prefactor = 0.53$ as we found in the effective theory.
Surprisingly, we find that for the given set of parameters, the separability condition holds extremely well 
(within machine precision) in the microscopic theory, which explains the numerical agreement between the two
results. Note that for
the same set of parameters, the indicated value of the coefficient $\Prefactor$ is close to the original
finding $\Prefactor \approx 1/2$ of Ref.~\onlinecite{baireuther2016scattering}.

\begin{figure}
\includegraphics[width=0.5\textwidth]{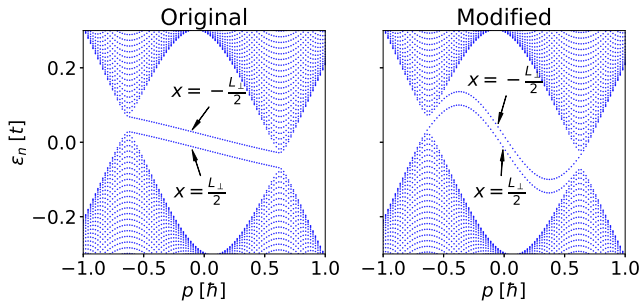}%
\caption{\label{fig:spectrum-original-deformed-boundary} Energy spectrum for the original (left) and the modified boundaries (right) of the microscopic Hamiltonian~\eqref{eq:Vazifeh-Franz-Hamiltonian}. In both cases, surface states (Fermi arcs) form continuous lines connecting the Weyl nodes, but the shape of a given line is sensitive to the structure of the boundary. The parameters of the Hamiltonian are taken from Sec.~\ref{sec:surface-microscopic-theory}, $p^y = 10^{-2} \hbar$, $B_0 = 0$. In order to illustrate the finite-size quantization of the energy bands, the number of lattice sites was decreased to 400.}
\end{figure}

The next step is to see whether the result is robust against microscopic deformation 
of the boundary. We have modified the boundary by rescaling the $\beta_0$
parameter, $\beta_0 \to 10 \beta_0$, at the boundary sites of the 1D
lattice. We have verified that such a modification does not affect the
parameters of the bulk effective theory~\eqref{eq:right-chiral-effective-Hamiltonian-VF}.
At the same time, the phase-shifts changed significantly,
$\Delta\phi^\pm_\textsc{r} = - \Delta\phi^\pm_\textsc{l} = \pm 2.3$. The
resulting drastic change of the Fermi-arc dispersion is illustrated in
Fig.~\ref{fig:spectrum-original-deformed-boundary}. Apart from the
quantitative modification of $\Delta\phi$, we have encountered a
qualitative change: the separability condition for the Fermi arcs
(Eq.~\eqref{eq:FA-separability}) is no longer satisfied. Thus, we conclude that
the separability is rather an \emph{accidental} property of the microscopic
theory~\eqref{eq:Vazifeh-Franz-Hamiltonian} with a specific boundary condition.
The numerical diagonalisation for the modified boundary leads to $\Prefactor = 1.05$,
which is quite different from the result for the original boundary.
By recalling the result $\Prefactor = 0.89$ of the effective theory from
Sec.~\ref{sec:surface-effective-theory} (for the same phaseshifts $\pm 2.3$),
we see that it is close to the finding within the microscopic theory,
although the agreement between the two approaches is not that good anymore.
This is a consequence of violating the separability condition and it makes the
prediction of the effective theory unreliable.

\section{Conclusion \label{sec:discussion}}
In this paper, we have considered the adiabatic chiral magnetic effect (CME) in a WSM sample having a
boundary. Generally, the contribution of the boundary to the CME current is on the same order
as the bulk contribution, in particular both are proportional to the
cross-sectional area of the sample. However we find that the boundary current can
dominate in a presence of a strong static background magnetic field. This is true if the theory possesses a certain symmetry or if the parameters of the effective Hamiltonian are fine-tuned in a certain way, as is discussed in Sec.~\ref{sec:bulk-effective-theory}.

We have found that there is no topological protection for such a boundary
current and in general it cannot even be determined from the parameters of a bulk low energy effective theory. However, under a certain assumption (separability of the Fermi arc energy,
Eq.~\eqref{eq:FA-separability}), the boundary current still can be expressed in terms of the
parameters of the bulk effective theory alone (by which we mean the
combination of the linearized Hamiltonian~\eqref{eq:linear-Hamiltonian-Weyl}
and the boundary conditions~\eqref{eq:boundary-condition-Weyl}). Such an expression can be found in Eq.~\eqref{eq:surface-current-separable}.

We have investigated the validity of the separability assumption in a
particular microscopic model~\eqref{eq:Vazifeh-Franz-Hamiltonian} used in
Refs.~\onlinecite{vazifeh2013electromagnetic} and~\onlinecite{baireuther2016scattering}. This model
accidentally has separable energy of the Fermi arcs, and the parameters of the Hamiltonian chosen in
Ref.~\onlinecite{baireuther2016scattering} were so that the abovementioned fine-tuning takes
place. This results in the surface current~\eqref{eq:surface-current-separable} being the
only source for the adiabatic CME. However, we see that a deformation of the
boundary layer in the model~\eqref{eq:Vazifeh-Franz-Hamiltonian} both breaks
the separability and makes Eq.~\eqref{eq:surface-current-separable} invalid.
In conclusion, the adiabatic CME current in a bounded Weyl semimetal system
is non-universal, but depends on the precise way one manufactures the
boundaries of the sample.

\begin{acknowledgments}
The authors are grateful to Paul Baireuther, Carlo 
Beenakker, Mikhail Katsnelson, and J\"{o}rg Schmalian for discussions and useful comments. This research was supported by the Foundation for Fundamental Research on Matter (FOM) and the Netherlands Organization for Scientific Research (NWO/OCW)  through the Delta ITP Consortium and by an ERC Synergy Grant.
\end{acknowledgments}

\appendix
\section{Relations between the parameters of the effective and the microscopic Hamiltonians}
\label{app:appendix}
The effective parameters entering Eq.~\eqref{eq:right-chiral-effective-Hamiltonian-VF} are expressed via the original parameters of the microscopic Hamiltonian~\eqref{eq:Vazifeh-Franz-Hamiltonian} as
\begin{equation}
  \label{eq:effective-parameters-1}
  v_\perp = 2\sqrt{\frac{\beta_z^2 - \beta_0^2}{\beta_z^2 - 4 \epsilonNodeR^2 }}
\end{equation}
\begin{equation}
  \label{eq:effective-parameters-2}
  v_\parallel = \frac{\sqrt{ \mathcal{K}}}{\beta_z^2 - 4\epsilonNodeR^2},
\end{equation}
\begin{equation}
  v_\mathbb{1} = -\frac{4\epsilonNodeR (1+M_0) \sin \pzWR + \beta_0 \beta_z \cos \pzWR}{\beta_z^2 - 4\epsilonNodeR^2},
\end{equation}
where
\begin{multline}
  \mathcal{K} = (4\epsilonNodeR(1+M_0)\sin \pzWR + \beta_0 \beta_z \cos \pzWR)^2 \\ + (\beta_z^2 - 4\epsilonNodeR^2)(4 (1+M_0)^2 \sin^2 \pzWR - \beta_0^2 \cos^2 \pzWR).
  \nonumber
\end{multline}
Here, $\pzWR$ is a positive solution of
\begin{equation}
  \left( \frac{\beta_0}{\beta_z} \sin \pzWR \right)^2 + 2 (1+M_0) \cos \pzWR = 2 + 2M_0 + M_0^2 - \frac{\beta_z^2 - \beta_0^2}{4},
\end{equation}
and the energy of right-chiral node is
\begin{equation}
  \epsilonNodeR = - \frac{\beta_0}{\beta_z} \sin \pzWR.
\end{equation}
(For simplicity, we have set the hopping parameter equal to 1, $t=1$.)

\section{Higher-order corrections to the linearized effective Hamiltonian}
\label{app:higher-order-corrections}
By 
using the minimal effective Hamiltonian~\eqref{eq:linear-Hamiltonian-Weyl}, we 
implicitly assume that the coupling to the magnetic field $B$ is captured 
completely by replacing the quasimomentum $\bm p = -i\hbar\bm\nabla$ with the 
operator $-i\hbar\bm\nabla - e \bm A /c$. However, since our particles have real 
spin $\bm s$ (as opposed to the pseudospin $\hbar\bm\sigma/2$), they are 
expected to have Zeeman coupling, which introduces the correction $\Delta 
\mathcal{H}_\text{eff} = - g \muBohr \bm s \bm B$, where $g$ is the $g$-factor 
and $\muBohr$ is the Bohr magneton. Then the energy gap between the two 
neighbouring Landau levels is of order of $v \hbar / \lB \sim e^2 / \lB$, 
according to Eq.~\eqref{eq:bulk-Landau-level-dispersion}, while the corrections 
coming from the Zeeman coupling are expected to be suppressed by an additional 
factor $\aNN / \lB$, which is small for realistic fields $B \lesssim 10^6 
\Gauss$ that can be reached in laboratories in the foreseeable future. Here we 
have estimated the typical velocity $v$ of an electron and the crystalline lattice spacing $\aNN$ to be of order of the corresponding atomic units, $v \sim 
e^2/\hbar$, $\aNN \sim \hbar^2 / m_e e^2$, which are not quite far 
from the results of the band-structure calculations for some 
WSMs~\cite{lee2015fermi} and the direct X-ray diffraction 
measurements~\cite{boller1963transposition,xu2015discovery}. This irrelevance of 
Zeeman coupling is similar to what happens in 
graphene~\cite{goerbig2011electronic}, although contrary to graphene, where the 
$g$-factor is not very far from the ``bare'' value $g=2$, see Ref.~\onlinecite{zhang2006landau} 
we can have much larger $g$, which can in principle alter our conclusion. (In 
the estimate above, we assumed $g \sim 1$. For a similar discussion regarding 
the importance of the Zeeman coupling to WSMs, see Ref.~\onlinecite{goswami2015optical}.) 
Moreover, in materials with strong spin-orbit coupling such as the transition-metal 
monopnictides that were the first experimentally discovered 
WSMs~\cite{hasan2017discovery}, additional terms in the effective theory are 
allowed. One such term is $C_b\bm B \cdot \bm b$, where $\bm b = \bm p_0(\text{right}) - \bm 
p_0(\text{left})$ is the momentum separation of the Weyl nodes. Also we neglect 
the higher-derivative corrections to the dispersion relation of a Weyl fermion, 
such as the quadratic term $\Delta \mathcal{H}_\text{eff} = C_\text{2} ~ (\bm p - 
\bm p_0)^2 / 2m_e$, where $m_e$ is the ``bare'' electron mass. However, 
a similar kind of dimensional analysis reveals that
in the absence of 
some ``anomalously'' large coupling constants $C_b$ and $C_2$, 
the resulting corrections are expected to be suppressed as well.

\bibliography{references-my.bib}
\end{document}